\documentclass{bmcart}

\usepackage{amsthm,amsmath}
\usepackage[utf8]{inputenc} 

 \usepackage[caption=false,font=normalsize,labelfont=sf,textfont=sf]{subfig}

\usepackage{tikz,pgfplots}
\usetikzlibrary{decorations.pathreplacing,plotmarks}
\pgfplotsset{compat=1.7}

\startlocaldefs

\endlocaldefs

\begin{document}

\begin{frontmatter}

\begin{fmbox}
\dochead{Research}


\title{Robust single- and multi-loudspeaker least-squares-based equalization for hearing devices}


\author[
  addressref={UOLSP},                   
  corref={UOLSP},                       
  noteref={n1},                        
  email={henning.schepker@uni-oldenburg.de}   
]{\inits{H.S.}\fnm{Henning} \snm{Schepker}}
\author[
  addressref={UOLMP},                   
  noteref={n2},                        
  email={florian.denk@uni-oldenburg.de}   
]{\inits{H.S.}\fnm{Florian} \snm{Denk}}
\author[
  addressref={UOLMP},                   
  email={birger.kollmeier@uni-oldenburg.de}   
]{\inits{H.S.}\fnm{Birger} \snm{Kollmeier}}
\author[
  addressref={UOLSP},                   
  corref={UOLSP},                       
  email={simon.doclo@uni-oldenburg.de}   
]{\inits{S.D.}\fnm{Simon} \snm{Doclo}}


\address[id=UOLSP]{
  \orgdiv{Signal Processing Group, Department of Medical Physics and Acoustics and Cluster of Excellence Hearing4all},             
  \orgname{University of Oldenburg},          
  \city{Oldenburg},                              
  \cny{Germany}                                    
}
\address[id=UOLMP]{
  \orgdiv{Medizinische Physik, Department of Medical Physics and Acoustics and Cluster of Excellence Hearing4all},             
  \orgname{University of Oldenburg},          
  \city{Oldenburg},                              
  \cny{Germany}                                    
}


\begin{artnotes}
\note[id=n1]{Currently with Starkey Hearing Technologies, Eden Prarie, Minnesota, United States.} \note[id=n2]{Currently with German Institute of Hearing Aids, L{\"u}beck, Germany.} 
\end{artnotes}

\end{fmbox}


\begin{abstractbox}

\begin{abstract} 
To improve the sound quality of hearing devices, equalization filters can be used that aim at achieving acoustic transparency, i.e., listening with the device in the ear is perceptually similar to the open ear. The equalization filter needs to ensure that the superposition of the equalized signal played by the device and the signal leaking through the device into the ear canal matches a processed version of the signal reaching the eardrum of the open ear. Depending on the processing delay of the hearing device, comb-filtering artifacts can occur due to this superposition, which may degrade the perceived sound quality. In this paper we propose a unified least-squares-based procedure to design single- and multi-loudspeaker equalization filters for hearing devices aiming at achieving acoustic transparency. To account for non-minimum phase components, we introduce a so-called acausality management. To reduce comb-filtering artifacts, we propose to use a frequency-dependent regularization. Experimental results using measured acoustic transfer functions from a multi-loudspeaker earpiece show that the proposed equalization filter design procedure enables to achieve robust acoustic transparency and reduces the impact of comb-filtering artifacts. A comparison between single- and multi-loudspeaker equalization shows that for both cases a robust equalization performance can be achieved for different desired open ear transfer functions.
\end{abstract}


\begin{keyword}
\kwd{sound pressure equalization}
\kwd{acoustic transparency}
\kwd{hearing device}
\kwd{multi-loudspeaker}
\end{keyword}


\end{abstractbox}
%

\end{frontmatter}




\section{Introduction}
Despite major improvements in hearing device technology in the past decades, the acceptance of hearing aids and assistive listening devices is still rather limited, partly due to a suboptimal sound quality \cite{Killion2004,Sockalingam2009}. This is most prominent in first-time users and users with normal hearing or mild-to-moderate hearing loss. While these users would benefit from advanced hearing device processing like noise reduction, dereverberation and dynamic range compression, they usually do not accept degradations of the sound quality. In order to improve the sound quality, equalization algorithms have been proposed that aim at achieving so-called acoustic transparency \cite{Denk2017,Hoffmann2013,Valimaki2015,Ramo2014,Liski2016}, i.e., listening with the device inserted in the ear achieves a similar perceptual impression as listening without the device inserted.

Generally, equalization algorithms for acoustic transparency aim at matching the sound pressure reaching the eardrum when the device is inserted in the ear (aided ear) with the sound pressure at the eardrum when the device is not inserted (open ear) \cite{Denk2017,Valimaki2015}. For the open ear, the sound pressure at the eardrum only consists of the direct sound. In contrast, for the aided ear the sound pressure at the eardrum consists of the superposition of the direct sound leaking into the (partially) occluded ear canal and the sound picked up by the microphone(s) of the device, processed and played back by the loudspeaker(s) of the device. Since the sound played back by the device is typically delayed compared to the direct sound, so-called comb-filtering effects frequently occur which may degrade the perceived sound quality \cite{Stone2008,Schepker2019,Schepker2020}. Several equalization algorithms for hearing devices have been proposed in the literature \cite{Denk2017,Hoffmann2013,Ramo2014,Gupta2019,Schepker2018,Denk2018a,Fabry2019}. However, often either the direct sound component was neglected in the equalization filter design, e.g., \cite{Hoffmann2013}, or electro-acoustic components were neglected, e.g., \cite{Ramo2014}. Additionally, it may be desirable to include knowledge about the advanced hearing device processing when designing the equalization filter, e.g., when a hearing loss needs to be compensated. However, including knowledge about advanced hearing device processing has often been neglected in previous research, e.g., \cite{Hoffmann2013,Ramo2014,Denk2018a,Fabry2019}. In this paper we propose to include information about both the direct sound component as well as the hearing device processing in the equalization filter design.

Equalization in hearing devices is commonly performed using a single loudspeaker \cite{Valimaki2015}, i.e., a single equalization filter is computed to match the sound pressure of the aided ear and the open ear. Computing this equalization filter usually requires the inversion of the (estimated) acoustic transfer function (ATF) between the hearing device loudspeaker and the eardrum. However, since this ATF typically has zeros inside and outside the unit circle, perfect inversion with a stable and causal filter cannot be achieved \cite{Kodrasi2014,Miyoshi1988}. Hence, approximate solutions are required to obtain a good equalization filter when using a single loudspeaker \cite{Denk2017,Hoffmann2013,Ramo2014,Denk2018a,Fabry2019}, e.g., equalizing only the minimum-phase component \cite{Ramo2014,Denk2017} or by including a so-called acausal delay \cite{Denk2018a,Fabry2019}. On the contrary, using multiple loudspeakers perfect equalization can be achieved when the conditions of the multiple-input/output inverse theorem (MINT) are satisfied \cite{Miyoshi1988}. Briefly, MINT states that perfect inversion of a multi-channel system can be achieved if all channels are co-prime, i.e., they do not share common zeros, and the equalization filters are of sufficient length. However, since multi-loudspeaker equalization using MINT is known to be very sensitive to small changes in the ATFs \cite{Radlovic2000}, regularization is commonly applied to increase the robustness \cite{Hikichi2007,Kodrasi2013} or other optimization criteria are considered \cite{Lim2014,Mertins2010}. Multi-loudspeaker equalization for acoustic transparency in hearing devices was considered in \cite{Schepker2018}, where the equalization filters were shown to exhibit common zeros, rendering the application of MINT difficult.

In this paper we propose a unified procedure to design an equalization filter that can be applied when using either a single loudspeaker or multiple loudspeakers to achieve acoustic transparency. The equalization filter is computed by minimizing a least-squares cost function, where we show that for the considered scenario the multi-loudspeaker system exhibits common zeros. Since these common zeros are, however, exactly known, we propose to exploit this knowledge and reformulate the optimization problem accordingly. In order to account for potential non-minimum phase components, we propose to incorporate an acausal delay in the filter computation, similarly as proposed for single-loudspeaker equalization in \cite{Denk2018a,Fabry2019}. Furthermore, to counteract comb-filtering effects we propose a frequency-dependent regularization term to reduce the hearing device playback when the leakage signal and the desired signal at the eardrum are of similar magnitude, similar as proposed for single-loudspeaker equalization in \cite{Denk2018a}. While regularization can also be used to increase the robustness of the equalization filters to unknown acoustic transfer functions, in this paper we propose to improve the robustness by considering multiple set of measurements in the optimization. Although some of these ideas were already presented in \cite{Schepker2018,Denk2018a}, the main objective of this paper is to present a unified procedure that can be used both for single-loudspeaker equalization and multi-loudspeaker equalization.

\begin{figure}[t]
	\centering
	\begin{tikzpicture}[scale=1.23]
		\node at (0,0) {\includegraphics[scale=0.615]{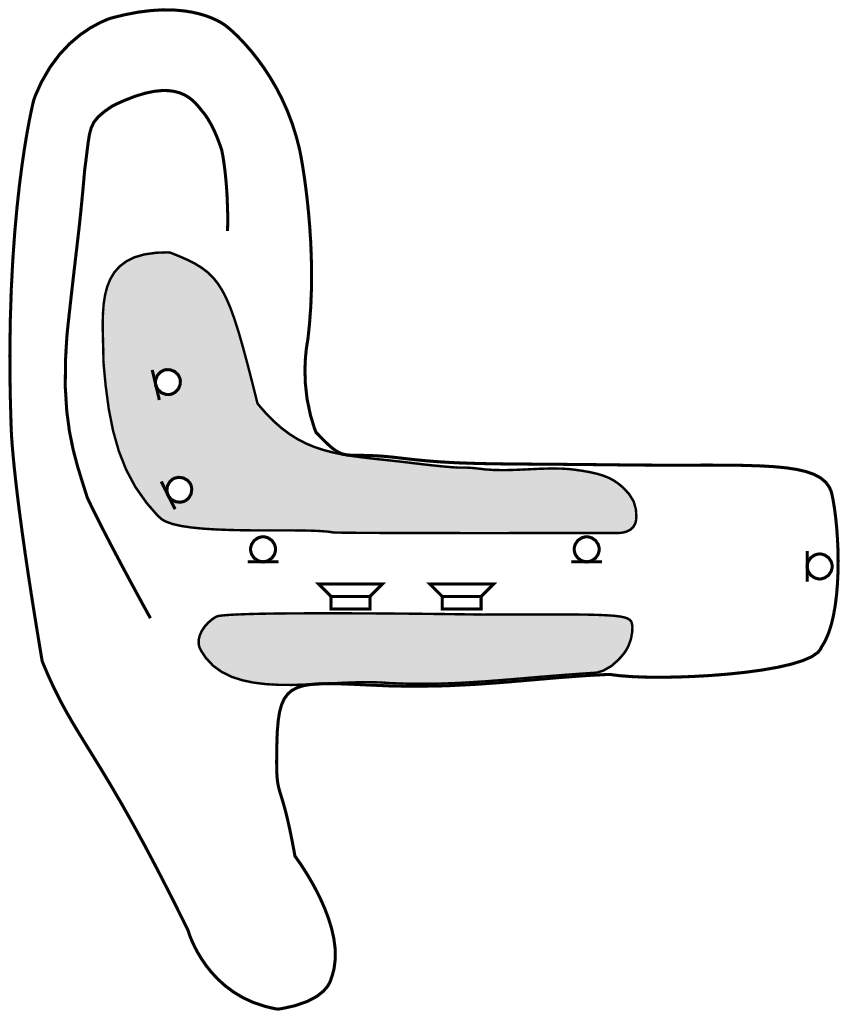}};
		\draw[thick,->] (-1.19,0.13) -- (0.5,1.05) -- (2.2,1.05) - | (3.85,0.0);
		\node at (0.2,0.6) {{\footnotesize $y[k]$}};
		\draw (2.2,-1.45) rectangle (3.2,-2.15);
		\node at (2.7,-1.8) {$\mathbf{A}(q)$};
		\draw[thick,->] (3.85,-0.75) |- (3.2,-1.8);
		\draw[thick,->] (2.20,-1.77) -| (0.20,-0.5);
		\draw[thick,->] (2.20,-1.83) -| (-0.36,-0.5);
		\draw[fill=white] (3.4,-0.75) rectangle (4.3,-0.0);
		\node at (3.85,-0.375) {{\footnotesize $G(q)$}};
		\node at (3.55,-1.25) {{\footnotesize $\tilde{u}[k]$}};
		\node[rotate=-90] at (2.35,-0.340) {{\footnotesize eardrum}};
		\node at (1.17,-1.6) {{\footnotesize $u_1[k]$}};
		\node at (-0,-2.1) {{\footnotesize $u_2[k]$}};
	\end{tikzpicture}
	\caption{Schematic overview of considered single-microphone multi-loudspeaker earpiece setup \cite{Denk2019}.}
	\label{fig:setup}
\end{figure}

Experimental results using measured ATFs from a multi-loudspeaker earpiece depicted in Figure \ref{fig:setup} show that the proposed single- and multi-loudspeaker equalization approach is able to achieve almost perfect equalization. Furthermore, we show that the equalization performance depends on the gain and the processing delay of the hearing device. By incorporating the frequency-dependent regularization, the effect of comb-filtering in the lower frequency region can be considerably reduced. Furthermore, robust equalization can be achieved by considering multiple sets of measurements when computing the equalization filter. A performance comparison between single- and multi-loudspeaker equalization shows that robust equalization can be achieved independent of the number of loudspeakers.

The remainder of this paper is organized as follows. In Section \ref{sec:scenario} we describe the considered hearing device setup. In Section \ref{sec:systemanalysis} we analyse the single-microphone-multiple-loudspeaker scenario with respect to the processing parameters of the hearing deivce. In Section \ref{sec:filtercomp} wepresent the robust single- and multi-loudspeaker equalization filter design procedure using a regularized least-squares cost function. In Section \ref{sec:experimentalEval} the proposed equalization filters are experimentally evaluated using either a single loudepaker or using multiple loudspeakers.

\section{Scenario and Problem Statement \label{sec:scenario}}
Consider a single-microphone-multi-loudspeaker hearing device with $N$ loudspeakers as depicted in Figure \ref{fig:simoframework}. 
\begin{figure}
 	\centering
 		\begin{scriptsize}
		\begin{tikzpicture}
		\draw (7.1,-2.45) rectangle (7.2,-2.05);
		\draw (7.1,-2.45) -- (6.95,-2.6) -- (6.95,-1.9) -- (7.1,-2.05);
		\draw[->] (6.7,-2.25) -- (5.0,-2.25);
		\draw[->] (6.7,-2.25) -| (6.0,-1.00);
		\node at (6.25,-2.45) {$s[k]$};
		\draw (6.7,1.25) circle (0.1);
		\draw (6.6,1.1) -- (6.6,1.4);
		\draw (6.8,1.25) -- (7.0,1.25);
		\node at (6.95,1.55) {eardrum};
		\node at (6.95,0.95) {$t_{aid}[k]$};
		\draw (3,1.95) rectangle (3.1,1.55);
		\draw (3.1,1.95) -- (3.25,2.1) -- (3.25,1.4) -- (3.1,1.55);
		\draw (3,0.95) rectangle (3.1,0.55);
		\draw (3.1,0.95) -- (3.25,1.1) -- (3.25,0.4) -- (3.1,0.55);
		\draw (1.9,0.60) rectangle (0.9,1.90);
		\node at (1.4,1.25) {$\mathbf{A}(q)$};
		\draw (1.9,0.75) -- (3.0,0.75);
		\draw (1.9,1.75) -- (3.0,1.75);
		\draw (-0.5,0) rectangle (0.5,-1.0);
		\node at (0,-0.5) {$G(q)$};
		\node at (0.35,0.5) {$\tilde{u}[k]$};
		\draw[->] (0,0) |- (0.9,1.25);
		\draw (3.15,-2.25) circle (0.1);
		\draw (3.25,-2.1) -- (3.25,-2.4);
		\draw (3.05,-2.25) -| (0.0,-1.0);
		\draw (5.45,0) rectangle (6.55,-1.0);
		\node at (6,-0.5) {$H_{occ}(q)$};
		\draw[->] (6.0,0) |- (6.35,1.15);
		\draw (4.0,0.6) rectangle (5.0,1.9);
		\node at (4.5,1.25) {$\mathbf{D}(q)$};
		\draw[->] (3.5,0.75) -- (4.00,0.75); 
		\draw[->] (3.5,1.75) -- (4.00,1.75); 
		\draw[->] (5.0,1.25) -- (6.35,1.25);
		\draw (4.0,-2.75) rectangle (5.0,-1.75);
		\node at (4.5,-2.25) {$H_m(q)$};
		\draw[->] (4.0,-2.25) -- (3.5,-2.25);
		\node at (2.6,-2.05) {$y[k]$};
		\node at (2.6,1.55) {$u_1[k]$};
		\node at (2.6,0.55) {$u_N[k]$};
		\node at (2.6,1.25) {$\vdots$};
	\end{tikzpicture}	
 		\end{scriptsize}
	\caption{Generic single-microphone multi-loudspeaker hearing device setup considered in this work.}
	\label{fig:simoframework}
\end{figure}
For simplicity we assume that all transfer functions are linear and time-invariant and that they can be modeled as polynomials in the variable $q$ \cite{Ljung1983}. We assume that the signal $y[k]$ picked up by the microphone of the hearing device is the signal emitted from a single directional sound source $s[k]$, i.e.,
\begin{align}
	y[k] = H_m(q)s[k], \label{eq:yk}
\end{align}
where $k$ denotes the discrete-time index and $H_m(q)$ denotes the ATF between the source and the microphone of the hearing device, i.e.,
\begin{align} 
H_m(q) = \sum_{i = 0}^{L_H-1}h_{m,i}q^{-i} = \mathbf{h}_m^T\mathbf{q},
\end{align}
 with  $\mathbf{q}$ a vector of delay elements and the $L_H$-dimensional impulse response (IR) vector $\mathbf{h}_m$ of $H_m(q)$ given by
\begin{align}
	\mathbf{h}_m = \begin{bmatrix} h_{m,0} & h_{m,1} & \dots & h_{m,L_H-1} \end{bmatrix}^T, \label{eq:vechm}
\end{align}
where $(\cdot)^T$ denotes the transpose operation. The microphone signal $y[k]$ is processed by the forward path $G(q)$ of the hearing device, which accounts for potential advanced processing and the processing delay of the device, yielding the intermediate signal $\tilde{u}[k]$, i.e.,
\begin{align}
	\tilde{u}[k] = G(q)y[k], \label{eq:tildeuk}
\end{align}
with the $L_G$-dimensional IR vector $\mathbf{g}$ of $G(q)$ defined similary as $\mathbf{h}_m$ in \eqref{eq:vechm}. The intermediate signal $\tilde{u}[k]$ is used as the input to $N$ equalization filters $A_n(q)$, $n=1,\dots,N$, yielding the $N$-dimensional loudspeaker signal vector $\mathbf{u}[k]$, i.e.,
\begin{align}
	\mathbf{u}[k] = \begin{bmatrix} A_1(q) & \dots & A_N(q) \end{bmatrix}^T \tilde{u}[k] = \mathbf{A}(q) \tilde{u}[k], \label{eq:vecuk}
\end{align}
with the $L_A$-dimensional equalization filter coefficient vector $\mathbf{a}_n$ of $A_n(q)$ given by
\begin{align}
	\mathbf{a}_n = \begin{bmatrix} a_{n,0} & a_{n,1} & \dots & a_{n,L_A-1} \end{bmatrix}^T.
\end{align}
Furthermore, we define the $NL_A$-dimensional vector of stacked equalization filter coefficient vectors as
\begin{align}
	\mathbf{a} = \begin{bmatrix} \mathbf{a}_1^T & \dots & \mathbf{a}_N^T \end{bmatrix}^T. \label{eq:veca}
\end{align}
For the aided ear, i.e., when the device is inserted and playing back the processed microphone signal, the signal $t_{aid}[k]$ at the eardrum of the listener is the superposition of the loudspeaker signals and the signal leaking into the (partially) occluded ear canal, i.e.,
\begin{align}
	t_{aid}[k] = \mathbf{D}^T(q)\mathbf{u}[k]+H_{occ}(q)s[k], \label{eq:tk}
\end{align}
where $H_{occ}(q)$ denotes the ATF between the source and the eardrum for the occluded ear, i.e., with the hearing device inserted and processing off, with $\mathbf{h}_{occ}$ the $L_H$-dimensional IR vector of $H_{occ}(q)$, defined similarly as $\mathbf{h}_m$ in \eqref{eq:vechm}. The $N$-dimensional vector $\mathbf{D}(q)$ contains the ATFs between the loudspeakers of the hearing device and the eardrum, i.e.,
\begin{align}
	\mathbf{D}(q) = \begin{bmatrix} D_1(q) & \dots & D_N(q) \end{bmatrix}^T,
\end{align}
with $\mathbf{d}_n$ the $L_D$-dimensional IR vector of $D_n(q)$ denoted by $\mathbf{d}_n$, defined similarly as  $\mathbf{h}_m$ in \eqref{eq:vechm}. 

The desired signal at the eardrum $t_{des}[k]$ is the signal reaching the eardrum of the listener when the device is not inserted (open ear), processed with the forward path of the device, i.e., 
\begin{align}
t_{des}[k] = G(q)\underbrace{H_{open}(q)s[k]}_{t_{open}[k]}, \label{eq:tdesk}
\end{align}
where $H_{open}(q)$ denotes the ATF between the source and the open ear, with $\mathbf{h}_{open}$ the $L_H$-dimensional IR vector of $H_{open}(q)$.
In order to achieve acoustic transparency, the goal is to obtain the equalization filters $a$ in \eqref{eq:veca} such that the signal $t_{aid}[k]$ in \eqref{eq:tk} is perceptually not distinguishable from the signal $t_{des}[k]$ in \eqref{eq:tdesk}, accounting for small variations in the ATFs, e.g., due to different positions of the hearing device in the ear.

%
\section{Transfer Function Analysis \label{sec:systemanalysis}}
In this section, we analyze the considered single-microphone multi-loudspeaker system in terms of its system transfer function between the source and the eardrum. For the aided ear, the system transfer function is obtained by combining \eqref{eq:yk}, \eqref{eq:tildeuk}, \eqref{eq:vecuk}, and \eqref{eq:tk}, leading to
\begin{align}
	H_{aid}(q) = \frac{t_{aid}[k]}{s[k]} = \mathbf{D}^T(q)\mathbf{A}(q)G(q)H_m(q) + H_{occ}(q). \label{eq:tkoversk}
\end{align}
Similarly, for the desired open ear transfer function, the system transfer function is obtained from \eqref{eq:tdesk} as
\begin{align}
	H_{des}(q) = \frac{t_{des}[k]}{s[k]} = G(q)H_{open}(q). \label{eq:tdeskoversk}
\end{align}
By equating \eqref{eq:tkoversk} and \eqref{eq:tdeskoversk} we observe that the optimal equalization filter needs to fulfill
\begin{align}
	G(q)H_m(q) \mathbf{D}^T(q)\mathbf{A}(q) = G(q)H_{open}(q) - H_{occ}(q), \label{eq:optimaleqfiltATF}
\end{align}
which corresponds to
\begin{align}
	\mathbf{D}^T(q)\mathbf{A}(q) = \frac{H_{open}(q)}{H_m(q)} - \frac{H_{occ}(q)}{H_m(q)} \frac{1}{G(q)}. \label{eq:optimaleqfiltRTF}
\end{align}
It should be noted that the optimal equalization filter in \eqref{eq:optimaleqfiltRTF} depends on the relative transfer functions (RTFs) $\frac{H_{open}(q)}{H_m(q)}$ and $\frac{H_{occ}(q)}{H_m(q)}$, i.e., the RTF between the eardrum and the microphone when the device is not inserted (open ear) and the RTF between the eardrum and the microphone when the device is inserted and switch off (occluded ear). Furthermore, the optimal filter in \eqref{eq:optimaleqfiltRTF} depends on the forward path $G(q)$. In order to analyze the dependency on the forward path, in the following we consider two extreme cases:
\begin{enumerate}
	\item Assuming no leakage component ($H_{occ}(q) = 0$), e.g., when the ear canal entrance is blocked completely by the device, the optimal equalization filter needs to fulfill 
	\begin{align}
		\mathbf{D}^T(q)\mathbf{A}(q) = \frac{H_{open}(q)}{H_m(q)},
	\end{align}
	such that $H_{aid}(q) = G(q)H_{open}(q)$.
	\item Assuming that $t_{des}[k]=0$, i.e., $H_{open}(q) = 0$, the optimal equalization filter aims at actively suppressing the leaking component at the eardrum, i.e., 
	\begin{align}
		\mathbf{D}^T(q)\mathbf{A}(q) = - \frac{H_{occ}(q)}{H_m(q)}\frac{1}{G(q)}, \label{eq:activesuppression}
	\end{align}
	such that $H_{aid}(q)=0$.
\end{enumerate}
The above analysis shows that for large forward path gains, equalization to the open ear ATF becomes more important than suppression of the leakage component, whereas for small forward path gains the leakage component dominates and needs to be actively suppressed. Furthermore, depending on the delay of $G(q)$, the equalized transfer function in \eqref{eq:optimaleqfiltRTF} may become increasingly acausal due to the term $\frac{1}{G(q)}$, which may impact the equalization performance. Additionally, the processing delay that can be allowed for these two cases is substantially different. For large forward path gains as in case 1, a large processing delay of a few milliseconds can be tolerated \cite{Stone2008}, while to achieve the desired active suppression in case 2, only very low processing delays of only a few microseconds can be tolerated \cite{Kuo1999}.

%
\section{Equalization filter design procedure \label{sec:filtercomp}}
In this section, we present a regularized least-squares-based design procedure to compute the equalization filter $\mathbf{A}(q)$. It should be noted that the same design procedure can be applied using either a single ($N=1$) or multiple ($N>1$) loudspeakers. In the following we assume knowledge of all required ATFs, e.g., by measurement. Alternatively some ATFs could also be estimated, e.g., an estimate of the open ear ATF between the source and the eardrum $H_{open}(q)$ could be obtained by an appropriate correction function of the ATF between the source and the microphone $H_m(q)$ \cite{Denk2018} or an estimate of the ATFs between the loudspeakers and the eardrum $\mathbf{D}(q)$ could be obtained using an in-ear microphone and an electro-acoustic model \cite{Vogl2017}. However, the investigation of such estimation procedures is beyond the scope of this paper. 

In Section \ref{subsec:optATFequal} we formulate the least-squares cost function for the equalization filter according to \eqref{eq:optimaleqfiltATF} and show that the transfer functions to be equalized share common zeros. Since these common zeros are exactly known, in Section \ref{subsec:optRTFequal} we exploit this knowledge and reformulate the least-squares cost function for the equalization filter according to \eqref{eq:optimaleqfiltRTF}, i.e., based on RTFs instead of ATFs.  To account for potential acausalities in the filter design, in Section \ref{subsec:acausalman} we incorporate an acausal delay in the optimization. In addition, in Section \ref{subsec:freqdepReg} we include a frequency-dependent regularization to reduce comb-filtering effects. Finally, in Section \ref{subsec:incrobustness} we include multiple measurements to increase the robustness of the equalization filters to variations, e.g., due to different positions of the hearing device in the ear.

\subsection{Optimal Equalization Filter using ATFs \label{subsec:optATFequal}}
The expression of the optimal equalization filter $\mathbf{A}(q)$ in \eqref{eq:optimaleqfiltATF} can be reformulated using matrix-vector notation as
\begin{align}
	\mathbf{C}\mathbf{a} = \mathbf{v}, \label{eq:matvecNotATF}
\end{align}
where $\mathbf{C}$ is an $(L_C + L_A - 1)\times NL_A$-dimensional matrix, with $L_C = L_G+L_H+L_D-2$, defined as
\begin{align}
	\mathbf{C} = \mathbf{G} \mathbf{H}_m \mathbf{D}, \label{eq:matC}
\end{align}
where $\mathbf{D}$ is the $(L_D+L_A-1)\times NL_A$-dimensional matrix of concatenated ($L_D+L_A-1) \times L_A$-dimensional convolution matrices $\mathbf{D}_n$ of the IR vector $\mathbf{d}_n$, i.e.,
\begin{align}
\mathbf{D} &= \begin{bmatrix} \mathbf{D}_1 & \dots & \mathbf{D}_N \end{bmatrix}, \label{eq:matD} \\
\mathbf{D}_n &= \begin{bmatrix} d_{n,0} & 0 & \dots & 0 \\
							   d_{n,1} & d_{n,0} & \ddots & \vdots \\
							   \vdots  & \ddots & \ddots & 0 \\
							   \vdots  & \ddots & \ddots & d_{n,0} \\
							   \vdots  & \ddots & \ddots & \vdots \\
							   d_{n,L_D-1}  & \ddots & \ddots & \vdots \\
							   0  & \ddots & \ddots & \vdots \\
							   \vdots  & \ddots & \ddots & \vdots \\
							   0	  & \dots & 0 & d_{n,L_D-1} \end{bmatrix},
\end{align}
$\mathbf{H}_m$ is the $(L_H + L_D + L_A - 2)\times (L_D + L_A -1)$-dimensional convolution matrix of the IR vector $\mathbf{h}_m$, and $\mathbf{G}$ is the $(L_C + L_A - 1)\times (L_H + L_D + L_A -2)$-dimensional convolution matrix of the IR vector $\mathbf{g}$.
Furthermore, $\mathbf{v}$ is the $(L_C + L_A - 1)$-dimensional vector of the desired equalization hearing device output
\begin{align}
	\mathbf{v} = \mathbf{G}\mathbf{\tilde{h}}_{open} - \mathbf{\tilde{h}}_{occ},
\end{align}
where $\mathbf{\tilde{h}}_{open}$ is the $(L_H + L_D + L_A - 2)$-dimensional zero-padded vector of the IR vector $\mathbf{h}_{open}$ and $\mathbf{\tilde{h}}_{occ}$ is the $(L_C + L_A - 1)$-dimensional zero-padded coefficient vector of the IR vector $\mathbf{h}_{occ}$.

The $NL_A$-dimensional equalization filter coefficient vector $\mathbf{a}$ is then obtained by minimizing the following least-squares cost function
\begin{align}
	J_{LS}^{atf} (\mathbf{a}) = \Vert \mathbf{C} \mathbf{a} - \mathbf{v} \Vert_2^2. \label{eq:JLS}
\end{align}

The optimal solution minimizing \eqref{eq:JLS} is equal to
\begin{align}
	\mathbf{a}_{LS}^{atf} = \mathbf{C}^\dag \mathbf{v}, \label{eq:LSsol}
\end{align}
where $(\cdot)^\dag$ denotes the pseudo-inverse of a matrix.

\subsection{Optimal Equalization Filter using RTFs \label{subsec:optRTFequal}}
Since the rows of the matrix $\mathbf{C}$ in \eqref{eq:matC} are linearly related by the matrix $\mathbf{G}\mathbf{H}_m$, the matrix $\mathbf{C}$ is not of full row-rank. In order to mitigate this rank deficiency\footnote{Note that regularization could also be used to mitigate this rank-deficiency. However, since we have perfect knowledge (in terms of the convolution matrices) of the common zeros, we decided to exploit this knowledge.}, we propose to left multiply both $\mathbf{C}$ and $\mathbf{v}$ by the pseudo-inverse of $\mathbf{G}\mathbf{H}_m$ (assumed to be of full column-rank), i.e.,
\begin{align}
	\mathbf{\tilde{C}} &=  (\mathbf{H}_m^T\mathbf{G}^T \mathbf{G} \mathbf{H}_m)^{-1} \mathbf{H}_m^T\mathbf{G}^T  \mathbf{C} = \mathbf{D}, \label{eq:matCtilde}\\
	\mathbf{\tilde{v}} &= (\mathbf{H}_m^T\mathbf{G}^T \mathbf{G} \mathbf{H}_m)^{-1} \mathbf{H}_m^T\mathbf{G}^T \mathbf{v}, \\
	&= (\mathbf{H}_m^T\mathbf{G}^T \mathbf{G} \mathbf{H}_m)^{-1} \mathbf{H}_m^T\mathbf{G}^T(\mathbf{G}\mathbf{\tilde{h}}_{open} - \mathbf{\tilde{h}}_{occ}),\label{eq:vecVtilde}
\end{align}
which is equivalent to writing \eqref{eq:optimaleqfiltRTF} using matrix-vector notation. It should be noted that $\mathbf{\tilde{v}}$ in \eqref{eq:matCtilde} represents an RTF, i.e., an infinite impulse response filter, which cannot be perfectly modeled using a finite impulse response filter and hence perfect equalization is not possible. Nevertheless, the least-squares cost function in \eqref{eq:JLS} can now be reformulated using $\mathbf{D}$ and $\mathbf{\tilde{v}}$ instead of $\mathbf{C}$ and $\mathbf{v}$, i.e.,
\begin{align}
	J_{LS}^{rtf} (\mathbf{a}) = \Vert \mathbf{D} \mathbf{a} - \mathbf{\tilde{v}} \Vert_2^2. \label{eq:JLSrtf}
\end{align}
However, since the ATFs between the loudspeakers and the eardrum $D(q)$ are likely to share near-common zeros due the close proximity of the loudspeakers in the considered hearing device (cf. Figure \ref{fig:setup}), the matrix inversion when using $\mathbf{D}$ in \eqref{eq:JLSrtf} is typically ill-conditioned. In order to mitigate this ill-conditioning, we add a regularization term to the least-squares cost function in \eqref{eq:JLSrtf} \cite{Hikichi2007,Kodrasi2013}, i.e.,
\begin{align}
	\boxed{J_{rLS}(\mathbf{a}) = \Vert \mathbf{D} \mathbf{a} - \mathbf{\tilde{v}} \Vert_2^2 + \lambda \Vert \mathbf{a} \Vert_2^2} \label{eq:JLSr}
\end{align}
where $\lambda$ is a real-valued non-negative regularization parameter. The optimal solution minimizing \eqref{eq:JLSr} is equal to
\begin{align}
	\boxed{\mathbf{a}_{rLS} = ( \mathbf{D}^T \mathbf{D} + \lambda\mathbf{I})^{-1} \mathbf{D}^T \mathbf{\tilde{v}}} \label{eq:LSrsol}
\end{align}
where $\mathbf{I}$ is the identity matrix and $\lambda$ is chosen to guarantee a numerically stable inversion of $\mathbf{D}^T \mathbf{D}$.

\subsection{Acausality Management \label{subsec:acausalman}}
While computing the equalization filter using \eqref{eq:LSrsol} may yield a reasonable performance, it has been shown in \cite{Denk2018a,Fabry2019} for single-loudspeaker equalization that allowing the filter design to account for acausalities can improve the equalization performance. This can be explained by the fact that accounting for acausalities allows for (partial) equalization of non-minimum phase components of the RTFs, and the inverse forward path gain $\frac{1}{G(q)}$ in \eqref{eq:optimaleqfiltRTF}. In the proposed single- and multi-loudspeaker equalization approach we, therefore, account for such potential non-minimum phase components, by delaying the transfer functions $H_{open}(q)$ and $H_{occ}(q)$ by $d_H$ samples, such that \eqref{eq:optimaleqfiltRTF} can be rewritten as
\begin{align}
	\mathbf{D}^T(q)\mathbf{A}(q) = \frac{H_{open}(q^{-d_H})}{H_m(q)} - \frac{H_{occ}(q^{-d_H})}{H_m(q)} \frac{1}{G(q)}. \label{eq:optimaleqfiltRTFacausal}
\end{align}
This corresponds to reformulating the cost function in \eqref{eq:JLSr} as
\begin{align}
	\boxed{J_{r\Delta LS}(\mathbf{a}) = \Vert \mathbf{D}_{\Delta} \mathbf{a} - \mathbf{\tilde{v}}_{\Delta} \Vert_2^2 + \lambda \Vert \mathbf{a} \Vert_2^2} \label{eq:JLSdeltar}
\end{align}
where $\mathbf{\tilde{v}}_{\Delta}$ is defined similarly as $\mathbf{\tilde{v}}$ in \eqref{eq:vecVtilde} but using the delayed open ear IR $\mathbf{\tilde{h}}_{open,\Delta}$ and the delayed occluded ear IR $\mathbf{\tilde{h}}_{occ,\Delta}$, i.e.
\begin{align}
	\mathbf{\tilde{v}}_{\Delta} &= (\mathbf{H}_{m,\Delta}^T\mathbf{G}_{\Delta}^T \mathbf{G}_{\Delta} \mathbf{H}_{m,\Delta})^{-1} \mathbf{H}_{m,\Delta}^T\mathbf{G}_{\Delta}^T (\mathbf{G}_{\Delta}\mathbf{\tilde{h}}_{open,\Delta} - \mathbf{\tilde{h}}_{occ,\Delta}), \label{eq:vecVtildeDelta}
\end{align}
with
\begin{align}
	\mathbf{\tilde{h}}_{open,\Delta} &= [ \underbrace{\begin{array}{ccc} 0 & \dots & 0 \end{array}}_{d_H} \,\,  \mathbf{\tilde{h}}_{open}^T \,\, ]^T, \\
	\mathbf{\tilde{h}}_{occ,\Delta} &= [ \underbrace{\begin{array}{ccc} 0 & \dots & 0 \end{array}}_{d_H} \,\,  \mathbf{\tilde{h}}_{occ}^T \,\, ]^T,
\end{align}
and defining the convolution matrices using zero-padded IRs, i.e.,
\begin{align}
	\mathbf{G}_{\Delta} &= \begin{bmatrix} \mathbf{G} & \boldsymbol{0} \\ \boldsymbol{0} & \boldsymbol{0} \end{bmatrix}, \\
	\mathbf{D}_{\Delta} &= \begin{bmatrix} \mathbf{D} & \boldsymbol{0} \\ \boldsymbol{0} & \boldsymbol{0} \end{bmatrix}, \label{eq:matTildeCdelta}\\
	\mathbf{H}_{m,\Delta} &= \begin{bmatrix} \mathbf{H}_{m} & \boldsymbol{0} \\ \boldsymbol{0} & \boldsymbol{0} \end{bmatrix}.
\end{align}
The optimal solution minimizing \eqref{eq:JLSdeltar} is equal to
\begin{align}
	\boxed{\mathbf{a}_{r\Delta LS} = ( \mathbf{D}_{\Delta}^T \mathbf{D}_{\Delta} + \lambda\mathbf{I})^{-1} \mathbf{D}_{\Delta}^T \mathbf{\tilde{v}}_{\Delta}} \label{eq:LSdeltarsol}
\end{align}
\subsection{Frequency-dependent Regularization \label{subsec:freqdepReg}}
Comb-filtering effects may occur due to constructive and destructive interference of the leakage component and the processed signal, which is delayed due to the processing delay of the hearing device. These effects are usually most pronounced in frequency regions where the leakage component $H_{occ}(q)s[k]$ is of similar level compared to the desired signal at the eardrum $t_{des}[k]$. Based on this observation, we propose to use a frequency-dependent regularization that aims at reducing comb-filtering effects by penalizing frequency regions where the magnitude of the leakage component is similar to the magnitude of desired signal, i.e., where
\begin{align}
	V(\omega_l) &=  \frac{\vert H_{occ}(\omega_l)\vert }{\vert H_{open}(\omega_l)G(\omega_l)\vert } \approx 1
\end{align}
where $\omega_l$ denotes the $l$-th angular frequency.

A frequency-dependent weighting factor is then computed using a zero mean logarithmic normal distribution with variance $\sigma^2 = \frac{\log 10}{20}\beta$, i.e., 
\begin{align}
W(\omega_l) = \frac{1}{\sqrt{2\pi}P(V(\omega_l))\sigma}e^{-\frac{1}{2} \left(\frac{\log P(V(\omega_l))}{\sigma}\right)^2 }, \label{eq:Womegal}
\end{align}
where the parameter $\beta$ enables to control the amount of regularization depending on the relative level of the leakage component and the desired signal and $P(\cdot)$ is a 1/6-octave smoothing with a rectangular smoothing window \cite{Hatziantoniou2000}. Using this weighting, we replace the frequency-independent regularization in \eqref{eq:JLSdeltar} with a frequency-dependent regularization, i.e.,
\begin{align}
	\boxed{J_{fr\Delta LS}(\mathbf{a})=\Vert \mathbf{D}_{\Delta} \mathbf{a} - \mathbf{\tilde{v}}_{\Delta} \Vert_2^2 + \lambda \Vert \mathbf{W} \mathbf{F} \mathbf{a} \Vert_2^2} \label{eq:JLSdeltarFreq}
\end{align}
where $\mathbf{F}$ is a $NL_{FFT}\times NL_A$-dimensional block-diagonal matrix consisting of $N$ $L_{FFT}\times L_A$-dimensional DFT matrices and $\mathbf{W}$ is a block-diagonal matix consisting of $N$ $L_{FFT}\times L_{FFT}$-dimensional diagonal matrices containing the weighting factors $W(\omega_l)$, $l=0,\dots,L_{FFT}-1$.
The optimal solution to \eqref{eq:JLSdeltarFreq} is equal to
\begin{align}
	\boxed{\mathbf{a}_{fr\Delta LS} = ( \mathbf{D}_{\Delta}^T \mathbf{D}_{\Delta} + \lambda\mathbf{F}^H\mathbf{W}^H\mathbf{W} \mathbf{F})^{-1} \mathbf{D}_{\Delta}^T \mathbf{\tilde{v}}_{\Delta}} \label{eq:LSdeltarFreqsol}
\end{align}
It should be noted that a similar frequency-dependent regularization was proposed in \cite{Denk2018a}. However, the regularization in \cite{Denk2018a} also limited the filter output when the desired signal at the eardrum $t_{des}[k]$ was much smaller than the leakage component, such that it is not applicable when active suppression of the leakage component is desired. On the contrary, the proposed weighting in \eqref{eq:Womegal} can also be used with small forward path gains, e.g., when the leakage component should be suppressed (cf. Section \ref{sec:systemanalysis}).

\subsection{Increased Robustness \label{subsec:incrobustness}}
While the frequency-dependent regularization allows to counteract comb-filtering effects, the obtained equalization filter may still be sensitive to variations in the ATFs, e.g., due to different positions of the hearing device in the ear. In order to increase the robustness to such variations, we propose to consider multiple sets of measured ATFs in the optimization, similarly as for single-loudspeaker equalization in \cite{Fabry2019}. 

Assuming that $I$ different sets of ATFs are available, we propose to extend the cost function in \eqref{eq:JLSdeltarFreq} as
\begin{align}
	\boxed{J_{mfr\Delta LS}(\mathbf{a}) = \sum_{i=1}^{I}\Vert \mathbf{D}_{\Delta,i} \mathbf{a} - \mathbf{\tilde{v}}_{\Delta,i} \Vert_2^2 + \lambda \Vert \mathbf{W} \mathbf{F}  \mathbf{a} \Vert_2^2} \label{eq:JLSmr}
\end{align}
where $\mathbf{\tilde{v}}_{\Delta,i}$ and $\mathbf{D}_{\Delta,i}$ are defined similarly as in \eqref{eq:vecVtildeDelta} and \eqref{eq:matTildeCdelta} for the $i$th set of ATFs, $i=1,\dots,I$. The optimal solution minimizing \eqref{eq:JLSmr} is equal to
\begin{align}	
	\boxed{\mathbf{a}_{mfr\Delta LS} = ( \mathbf{\bar{D}}_{\Delta}^T \mathbf{\bar{D}}_{\Delta} + \lambda\mathbf{F}^H \mathbf{W}^H \mathbf{W} \mathbf{F})^{-1} \mathbf{\bar{D}}_{\Delta}^T \mathbf{\bar{v}}_{\Delta}} \label{eq:LSmrsol}
\end{align}
with $\mathbf{\bar{D}}_{\Delta}$ the $I(L_D+L_A-1)\times NL_A$-dimensional matrix of stacked matrices $\mathbf{D}_{\Delta,i}$ and $\mathbf{\bar{v}}_{\Delta}$ the $I(L_D+L_A-1)$-dimensional vector of stacked vectors $\mathbf{\tilde{v}}_{\Delta,i}$, i.e.,
\begin{align}
	\mathbf{\bar{D}}_{\Delta} &= \begin{bmatrix} \mathbf{D}_{\Delta,1}^T & \dots & \mathbf{D}_{\Delta,I}^T\end{bmatrix}^T, \\
	\mathbf{\bar{v}}_{\Delta} &= \begin{bmatrix} \mathbf{\tilde{v}}_{\Delta,1}^T & \dots & \mathbf{\tilde{v}}_{\Delta,I}^T\end{bmatrix}^T.
\end{align}
The equalization filter in \eqref{eq:LSmrsol} is optimal in the mean across the ATFs considered in the optimization and thus is expected to be more robust to frequently occuring variations in the ATFs of the hearing device.

%
\section{Experimental Evaluation \label{sec:experimentalEval}}
In this section, we evaluate the performance of the proposed equalization design procedure, using a single loudspeaker ($N=1$) and using multiple loudspeaker ($N=2$). After introducing the considered setup and performance measures in Section \ref{subsec:setup}, we perform four different experiments. In Section \ref{subsec:res_acausal} we evaluate the impact of the acausality management. In Section \ref{subsec:res_reg} we investigate the impact of the frequency-dependent regularization. In Section \ref{subsec:res_robust} we investigate the robustness against unknown ATFs due to resinsertion of the hearing device in the ear. In Section \ref{subsec:res_gain} we evaluate the influence of different forward path gains on the equalization performance.

\subsection{Setup and Performance Measures \label{subsec:setup}}
All required ATFs were measured for the earpiece depicted in Figure \ref{fig:setup} (see also \cite{Denk2019},\cite{Denk2020subm}), which was inserted into the left ear of a GRAS 45BB-12 KEMAR Head \& Torso with low-noise ear simulators. It should be noted here, that this earpiece consist of four microphones and two loudspeakers. For the present evaluation we only used the microphone located on the outside close to the vent. The IRs of the ATFs were sampled at $f_s=16000$\,Hz and truncated to length $L_H = 130$ for the ATFs between the source and the earpiece and the eardrum and $L_D = 100$ for the ATF between the loudspeakers of the earpiece and the eardrum. Measurements were performed in an anechoic chamber with a distance of approximately $2.3$\,m between the frontal source and the dummy head. Each measurement was performed $I=5$ times after reinserting the earpiece to investigate reinsertion varibility. The forward path was set to $G(q) = 10^{G_0/20}q^{-d_G}$ with $G_0$ a broadband gain in dB and $d_G$ a delay in samples. Different broadband gains and delays were considered in the experiments.

To analyze the performance of the proposed equalization design procedure, we use the magnitude response of the aided ear transfer function $H_{aid}(q)$ in \eqref{eq:tkoversk} and the magnitude response of the desired open ear transfer function $H_{des}(q)$ in \eqref{eq:tdeskoversk}. To quantify the differences between both magnitude responses, we use a perceptually motivated auditory spectral distance, i.e.,
\begin{align}
	\Delta H_{aud} = \sum_{\omega_l = \omega_{low}}^{\omega_{up}} F(\omega_l) \Big\vert 10\log_{10}\frac{\vert H_{aid}(\omega_l)\vert^2}{\vert H_{des}(\omega_l)\vert^2} \Big\vert, \label{eq:metricDeltaH}
\end{align}
where $\omega_{low}$ and $\omega_{up}$ correspond to 200\,Hz and 8000\,Hz, respectively, and $F(\omega_l)$ is a frequency-dependent weighting function. To counteract over-representation of high frequencies, we have used the normalized inverse of the frequency-dependent equivalent rectangular bandwidth \cite{Glasberg1990} as weighting function, i.e.,
\begin{align}
	F(\omega_l) &= \frac{c}{24.7(4.37 \frac{\omega_l}{2\pi} + 1)},
\end{align}
where $c$ is a constant to ensure that the summation of the weighting function over the considered frequency range is equal to one.

In all experiments the equalization filter was computed using a filter length of $L_A = 99$, which is the optimal filter length for $N=2$. Note that for $N=1$ the optimal filter length of $L_A = \infty$ is obviously not realizable and does not guarantee perfect equalization.

\subsection{Experiment 1: Acausality management \label{subsec:res_acausal}}
In the first experiment, we investigate the impact of the acausality management proposed in Section \ref{subsec:acausalman}. For different values of the introduced acausal delay $d_H$, we computed the equalization filter using \eqref{eq:LSdeltarsol} for $N=1$ and $N=2$ loudspeakers, using a small regularization parameter $\lambda=10^{-8}$ to avoid numerical inversion problems. We used a broadband gain of $G_0 = 0$\,dB and a hearing device delay of either $d_G=1$ or $d_G=96$, corresponding to a delay of 0.0625\,ms and 6\,ms, respectively, which is well within the range of typical delays for commercial hearing devices with transparency features \cite{Denk2020}.  The same ATFs were used for computing the equalization filter and for evaluating its performance. Note that the sensitivity to unknown ATFs will be investigated in Experiment 3 (cf. Section \ref{subsec:res_robust}).

\begin{figure*}
	\centering
	\subfloat[$d_G=1$.\label{fig:exp1_n1_freq_dG1}]{\includegraphics[scale=0.70]{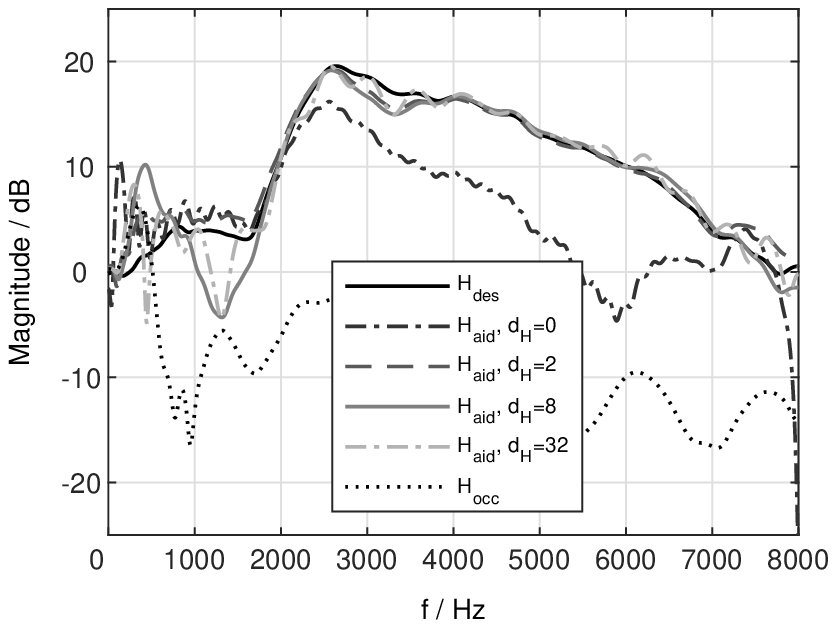}}\hfill
	\subfloat[$d_G=96$.\label{fig:exp1_n1_freq_dG96}]{\includegraphics[scale=0.70]{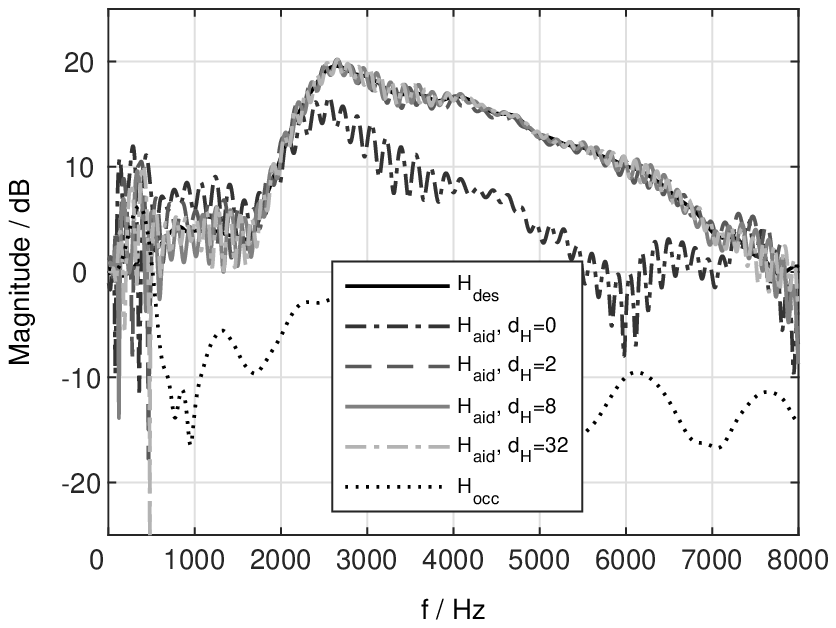}}
	\caption{Magnitude responses of the occluded ear transfer function $H_{occ}(q)$, the aided ear transfer function $H_{aid}(q)$ and the desired open ear transfer function $H_{des}(q)$ for different values of $d_H$ using $N=1$ loudspeaker, and $G_0=0$\,dB for (a) $d_G=1$ and (b) $d_G=96$.}	
\end{figure*}

For a hearing device delay of $d_G=1$ and $N=1$ loudspeaker, Figure \ref{fig:exp1_n1_freq_dG1} shows magnitude responses of the aided ear transfer function for different values of the acausal delay $d_H$  as well as the desired open ear transfer function and the occluded ear transfer function. As can be observerd, using no acausality management ($d_H=0$) leads to strong deviations of the aided ear transfer function from the desired open ear transfer function. By introducing an acausal delay ($d_H>0$), a better match between both transfer functions can be achieved. This is in line with results observed in \cite{Denk2018a,Fabry2019}. It should be noted that using a larger acausal delay may result in comb-filtering effects, in particular in frequency regions where the the occluded ear transfer function $H_{occ}(q)$ and the desired open ear transfer function $H_{des}(q)$ are of similar magnitude (here the frequency region below approximately 500Hz). In order to investigate the impact of the acausal delay for a larger hearing device delay, Figure \ref{fig:exp1_n1_freq_dG96} depicts the magnitude responses for $d_G=96$ and $N=1$ loudspeaker. As can be observed, comb-filtering effects now occur for all aided ear transfer functions. In addition to the comb-filtering effects, again strong deviations between the aided ear transfer function and the desired open ear transfer function occur for $d_H=0$, while a better match is obtained for $d_H>0$. Comparing the results for $d_G=1$ and $d_G=96$, despite the more pronounced comb-filtering effects for $d_G = 96$ only a small impact of the hearing device delay is observed, demonstrating that when considering single-loudspeaker equalization an acausal delay with $d_H \geq 1$ is crucial. In the following experiments the optimal value for $d_H$ will be determined.

\begin{figure*}
	\centering
	\subfloat[$d_G=1$.\label{fig:exp1_n2_freq_dG1}]{\includegraphics[scale=0.7]{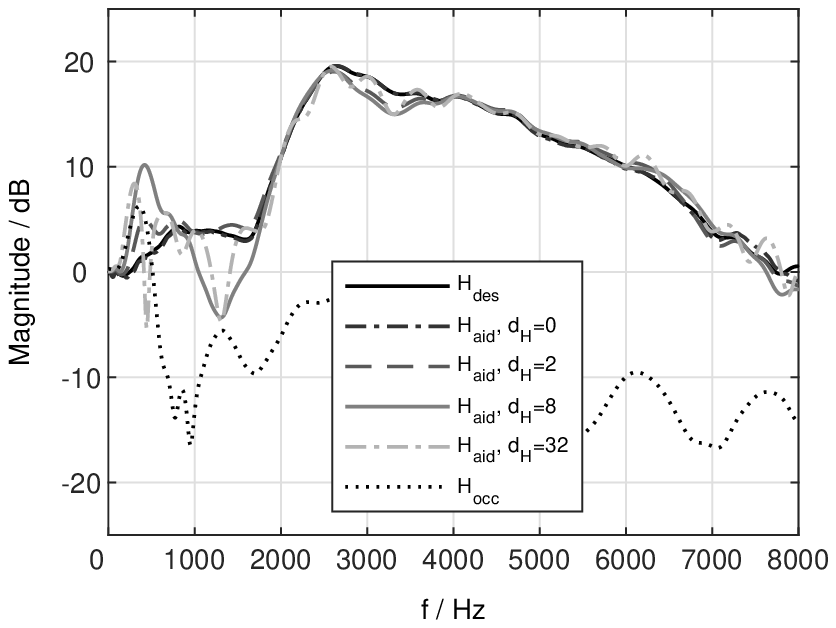}}\hfill
	\subfloat[$d_G=96$.\label{fig:exp1_n2_freq_dG96}]{\includegraphics[scale=0.7]{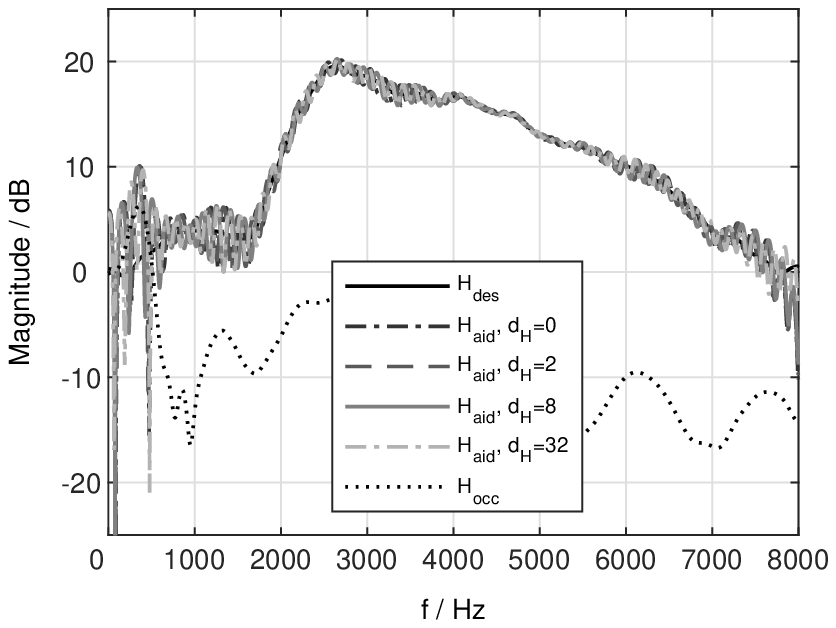}}
	\caption{Magnitude responses of the occluded ear transfer function $H_{occ}(q)$, the aided ear transfer function $H_{aid}(q)$ and the desired open ear transfer function $H_{des}(q)$ for different values of $d_H$ using $N=2$ loudspeaker, and $G_0=0$\,dB for (a) $d_G=1$ and (b) $d_G=96$.}	
\end{figure*}

For $N=2$ loudspeakers, Figure \ref{fig:exp1_n2_freq_dG1} and \ref{fig:exp1_n2_freq_dG96} show magnitude responses of the aided ear transfer function for different values of the acausal delay $d_H$ as well as the desired open ear transfer function and the occluded ear transfer function. In contrast to using $N=1$ loudspeaker, introducing an acausal delay ($d_H\geq 1$) does not yield a benefit compared to using no acausality management ($d_H=0$), but even leads to some deviations from the desired open ear transfer function in the lower frequencies due to comb-filtering effects. This can be explained by the fact that allowing for some acausality in a single-loudspeaker system makes it easier to equalize a non-minimum phase system, while for a multi-loudspeaker system a non-minimum phase system can be perfectly equalized without additional delays in case the MINT conditions are satisfied \cite{Miyoshi1988}. 

In order to investigate the impact of the acausal delay for a larger hearing device delay, Figure \ref{fig:exp1_n2_freq_dG96} depicts the magnitude responses for $d_G=96$ and $N=2$ loudspeaker. As can be observed, comb-filtering effects now occur for all aided ear transfer functions. Comparing the results for $d_G=1$ and $d_G=96$, despite the more pronounced comb-filtering effects for $d_G=96$, only a small impact of the hearing device delay is observed. These results demonstrate that when considering multi-loudspeaker equalization, an acausal delay is generally not necessary. However, as will be shown in the next experiment a larger $d_H$ may be beneficial.

\subsection{Experiment 2: Influence of regularization \label{subsec:res_reg}}
In the second experiment, we investigate the impact of the frequency-dependent regularization proposed in Section \ref{subsec:freqdepReg}. We will analyze the performance of the equalization filters computed for different values of the trade-off parameter $\lambda$ in \eqref{eq:LSdeltarFreqsol} and the control parameter $\beta$ in \eqref{eq:Womegal}. In this experiment we used a broadband gain of $G_0=0$\,dB and a hearing device delay of $d_G=96$. If not mentioned otherwise, we used an acausal delay of $d_H=32$ (this optimal value will be determed later in this section).

\begin{figure*}
	\centering
	\subfloat[$N=1$.\label{fig:exp2_n1_freq}]{\includegraphics[scale=0.7]{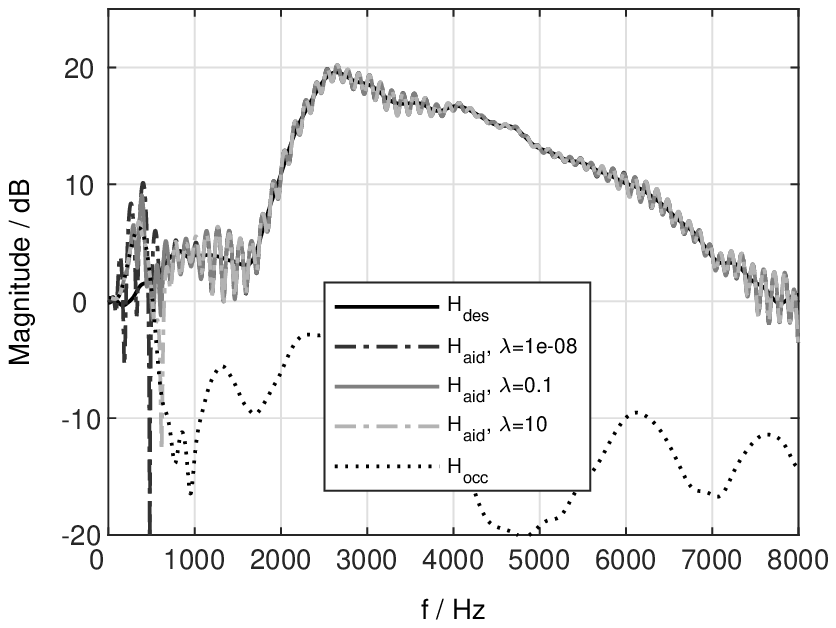}} \hfill
	\subfloat[$N=2$.\label{fig:exp2_n2_freq}]{\includegraphics[scale=0.7]{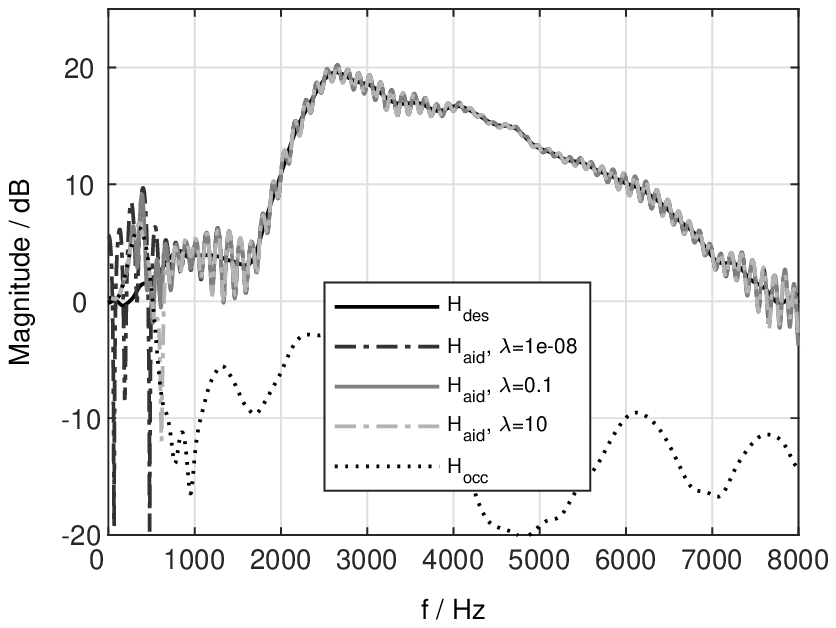}} \\
	\subfloat[Zoom $N=1$.\label{fig:exp2_n1_freq_zoom}]{\includegraphics[scale=0.7]{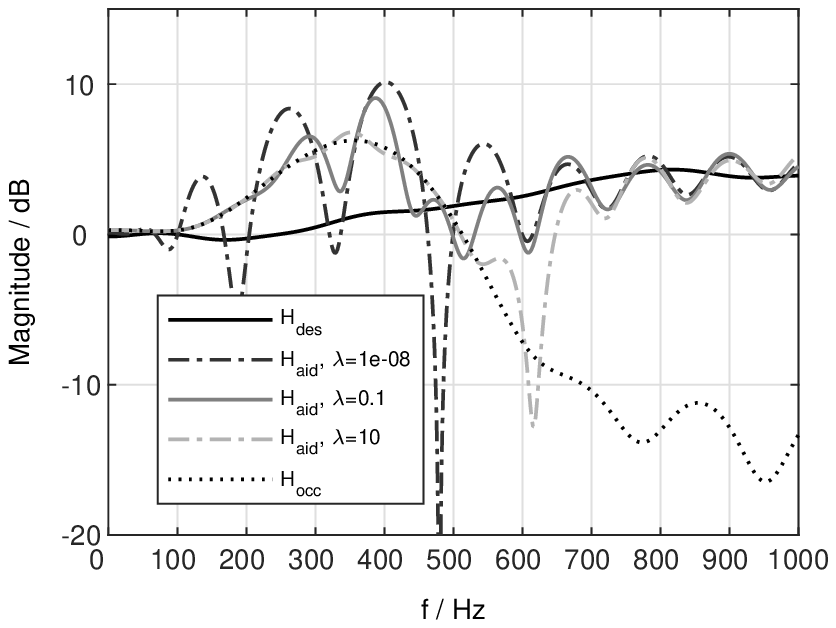}} \hfill
	\subfloat[Zoom $N=2$.\label{fig:exp2_n2_freq_zoom}]{\includegraphics[scale=0.7]{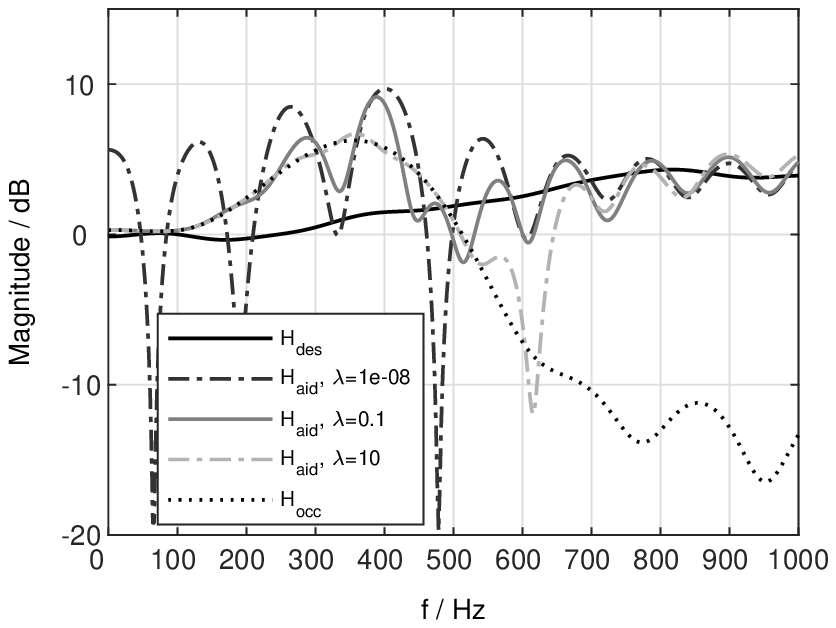}}
	\caption{Magnitude responses of the occluded ear transfer function $H_{occ}(q)$, the desired open ear transfer function $H_{des}(q)$ and the aided ear transfer function $H_{aid}(q)$ for different values of $\lambda$ for $N=1$ and $N=2$ loudspeakers. ($\beta=1$, $d_G=96$ and $G_0=0$\,dB)}
\end{figure*}

For $N=1$ loudspeakers, Figure \ref{fig:exp2_n1_freq} shows magnitude responses of the aided ear transfer function for different values of $\lambda$ and $\beta=1$ as well as the desired open ear transfer function and the occluded ear transfer function. As can be observed, for high frequencies no major differences can be observed for the different considered values of $\lambda$, while differences are visible in the lower frequencies especially for $f\leq 500$\,Hz, which is even clearer in the zoomed in portion in Figure \ref{fig:exp2_n1_freq_zoom}. This is due to the fact that regularization is mostly affecting frequency regions where the occluded ear transfer function $H_{occ}(q)$ and the desired open ear transfer function $H_{des}(q)$ are of similar magnitude. Therefore, in the following we will focus on the frequency region below 1\,kHz to assess the impact of the regularization parameter $\lambda$ and the control parameter $\beta$. As can be observed in Figure \ref{fig:exp2_n1_freq_zoom}, increasing $\lambda$ reduces undesirable comb-filtering effects but increases the similarity between the aided ear transfer function and the occluded ear transfer function. For example for the largest considered value of $\lambda=10$ no visible comb-filtering artifacts occur, but larger deviations between the aided ear transfer function and the desired open ear transfer function occur for frequencies between 500\,Hz and 700\,Hz compared to the smaller values of $\lambda$.  In general, the parameter $\lambda$ introduces a trade-off between a reduction of comb-filtering artifacts in the lower frequencies and a good equalization performance in frequency regions where the magnitude responses of the occluded ear transfer function and the desired open ear transfer function begin to deviate. 

\begin{figure*}
	\centering
	\subfloat[$N=1$.\label{fig:exp2_n1_dHs}]{\includegraphics[scale=0.7]{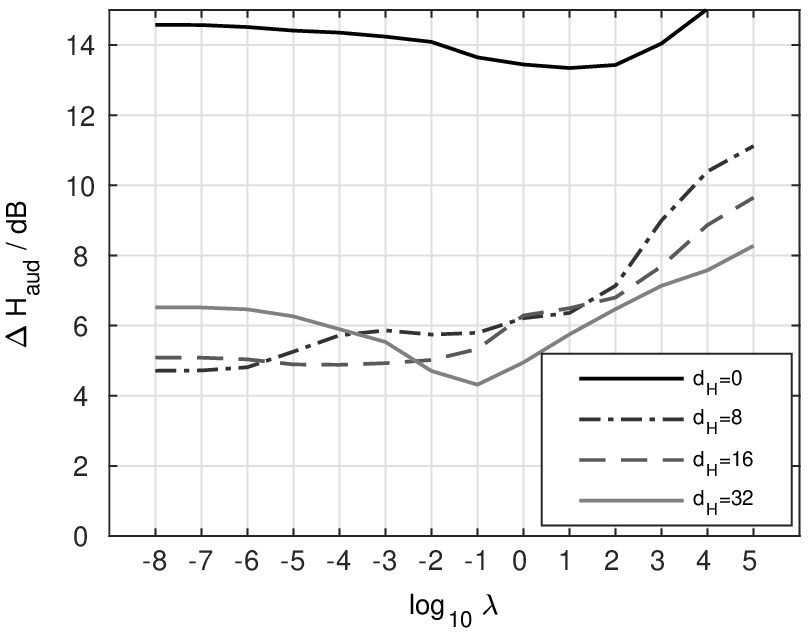}} \hfill
	\subfloat[$N=2$.\label{fig:exp2_n2_dHs}]{\includegraphics[scale=0.7]{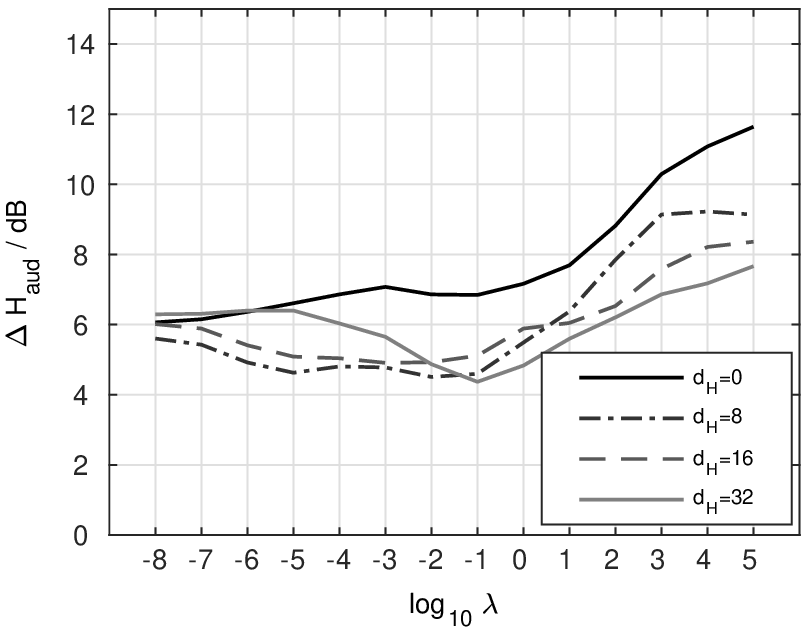}}
	\caption{Auditory spectral distance as a function of the regularization parameter $\lambda$ for different acausal delays $d_H$ (a) $N=1$ loudspeaker and (b) $N=2$ loudspeakers ($\beta=1$, $G_0=0$\,dB, $d_G=96$).}
\end{figure*}

\begin{figure*}
	\centering
	\subfloat[$N=1$.\label{fig:exp2_n1_beta}]{\includegraphics[scale=0.7]{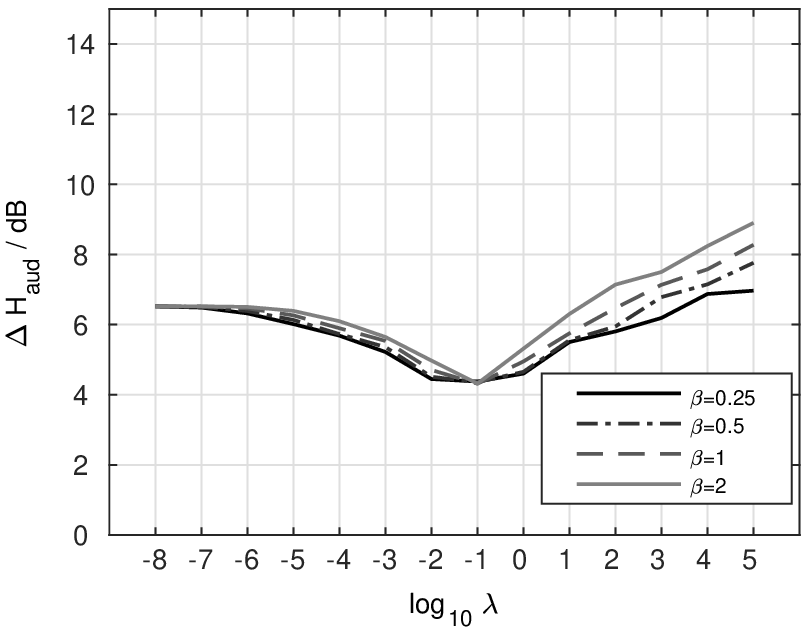}} \hfill
	\subfloat[$N=2$.\label{fig:exp2_n2_beta}]{\includegraphics[scale=0.7]{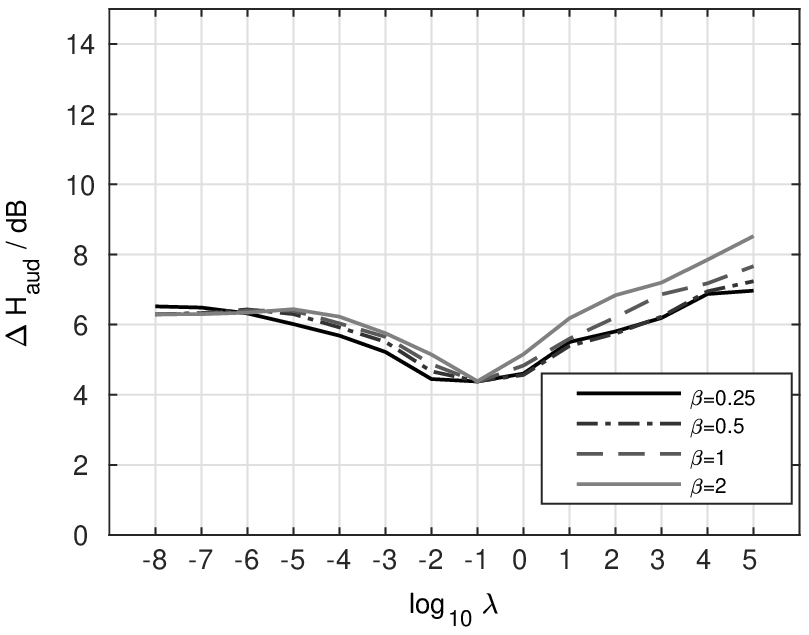}}
	\caption{Auditory spectral distance as a function of the regularization parameter $\lambda$ for different values of control parameter $\beta$ (a) $N=1$ loudspeaker and (b) $N=2$ loudspeakers ($d_H=32$, $G_0=0$\,dB, $d_G=96$).}
\end{figure*}

In order to investigate a potential interaction between the acausal delay $d_H$ and the regularization parameter $\lambda$, Figure \ref{fig:exp2_n1_dHs} depicts the auditory spectral distance $\Delta H_{aud}$ in \eqref{eq:metricDeltaH} as a function of $\lambda$ for different values of $d_H$ and $\beta=1$. In general, increasing the regularization parameter results in a larger auditory spectral distance. The proposed frequency-dependent regularization yields the lowest auditory spectral distance for $d_H=32$ and $\lambda = 0.1$. To investigate the impact of the control parameter $\beta$, Figure \ref{fig:exp2_n1_beta} depicts the auditory spectral distance as a function of $\lambda$ for different values of $\beta$ using $d_H=32$. As can be observed, the auditory spectral distance generally increases with increasing $\beta$. The lowest auditory spectral distance is obtained for $\beta =1$ and $\lambda=0.1$. These results show that when using the proposed approach with a single loudspeaker and a delay of $d_G=96$, using $\lambda = 0.1$ and $\beta=1$ are reasonable and allows to reduce comb-filtering effects in the lower frequency region while maintaining accurate equalization results.


For $N=2$ loudspeakers, Figure \ref{fig:exp2_n2_freq}, shows magnitude responses of the aided ear transfer function for different values of $\lambda$ and $\beta=1$, as well as the desired open ear transfer function and the occluded ear transfer function. Similarly as for $N=1$, for high frequencies no major differences can be observed for the different considered values of $\lambda$, while differences are visible in the lower frequencies, e.g., especially for $f\leq 1$\,kHz (cf. Figure \ref{fig:exp2_n2_freq_zoom}). Again, this is due to the fact that regularization is mostly affecting frequency regions where the occluded ear transfer function $H_{occ}(q)$ and the desired open ear transfer function $H_{des}(q)$ are of similar magnitude. Similarly as for $N=1$, Figure \ref{fig:exp2_n2_dHs} and Figure \ref{fig:exp2_n2_beta} show the auditory spectral distance as a function of $\lambda$ for different values of $d_H$ and $\beta$ when using $N=2$ loudspeakers. It can be observed that the lowest auditory spectral distance is obtained for the same parameters as for $N=1$, i.e., $d_H=32$, $\lambda=0.1$, and $\beta =1$. We will hence use these parameter values in the following two experiments.

\subsection{Experiment 3: Robustness against unknown ATFs \label{subsec:res_robust}}
While in the previous experiments same acoustic ATFs were used for computing and evaluating the performance of the equalization filter, in this experiment we investigate the impact of unknown ATFs on the performance of the equalization filter. To this end, we use five different sets of measured ATFs obtained after reinserting the earpiece into the ear of the dummy head and compute the equalization filter using the cost function defined in \eqref{eq:LSmrsol} with $I=4$ sets of ATFs. We evaluate the performance using the fifth set of ATFs that was not used for the computation of the equalization filter. This procedure is repeated for each of the five available sets of measurements, i.e., we use a leave-one-out cross-validation approach. In this experiment we used a broadband gain of $G_0=0$\,dB and hearing device delay of $d_G=96$.

\begin{figure*}
	\centering
	\subfloat[$N=1$.\label{fig:exp3_n1_robust}]{\includegraphics[scale=0.7]{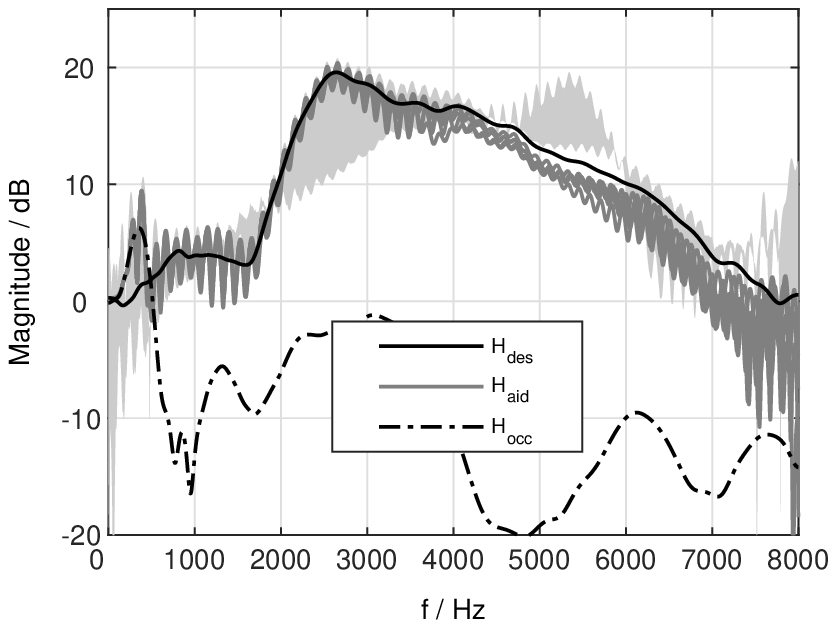}}
	\hfill
	\subfloat[$N=2$.\label{fig:exp3_n2_robust}]{\includegraphics[scale=0.7]{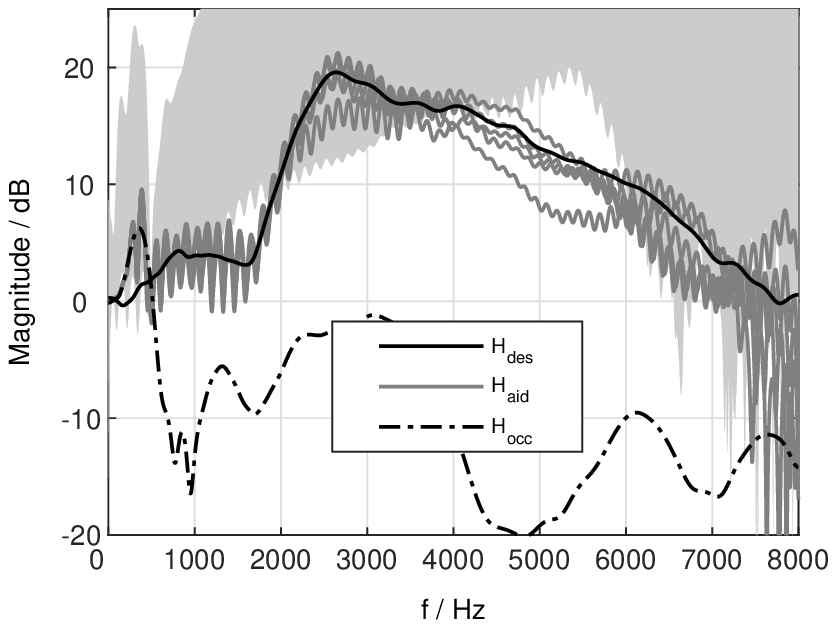}}
	\caption{Magnitude responses of the occluded ear transfer function $H_{occ}(q)$, the desired open ear transfer function $H_{des}(q)$ and the aided ear transfer function $H_{aid}(q)$ for the robust optimization for (a) $N=1$ and (b) $N=2$ ($\lambda=0.1$, $d_G=96$, $d_H=32$, $\beta=1$). The light grey shaded area shows the range of the results based on optimization using a single set of measurements.}
	\label{fig:exp3_robust}
\end{figure*}

Figure \ref{fig:exp3_robust} shows the magnitude responses of the aided ear transfer function for $N=1$ and $N=2$ loudspeakers, respectively, as well as the desired open ear transfer function and the occluded ear transfer function. For both single- and multiple-loudspeaker equalization it can be observed that the results obtained by using multiple sets of measurements to compute the equalization filter (grey curves) are much closer to the desired open ear transfer function than the range of results obtained using only a single set of measurements to compute the equalization filter (light grey shaded area in the background). This is particularly the case for multi-loudspeaker equalization, where huge deviations occur for unknown ATFs when using only a single set of measurements to compute the equalization filter. Comparing the results for $N=1$ and $N=2$, in general a slightly better approximation of the desired open ear transfer function is achieved using $N=1$, especially in the frequency range from 3500Hz to 6000Hz. These results demonstrate that both using a single loudspeaker as well as multiple loudspeakers a robust equalization can be achieved when considering multiple sets of measurements in the filter optimization, where single-loudspeaker equalization is slightly more robust than multi-loudspeaker equalization.

\subsection{Experiment 4: Influence of forward path gain \label{subsec:res_gain}}
While in the previous experiments we used a forward path gain of $G_0 = 0$\,dB, in practice also larger gains are obviously relevant. Therefore, in this experiment we investigate the impact of the forward path gain on the performance of the equalization filter. To this end, we consider 3 different broadband gains, i.e., $G_0 = 0$\,dB, $G_0 = 10$\,dB and $G_0 = 20$\,dB. Similarly as in Experiment 3, for each considered forward path gain we compute 5 different equalization filters  using $I=4$ sets of measured ATFs and use the fifth set of ATFs for evaluation in a leave-one-out cross-validation approach. We use the same parameter settings as in Experiment 3, i.e., $d_G = 96$, $d_H=32$, $\lambda=0.1$, and $\beta=1$.

\begin{figure*}
	\centering
	\subfloat[$N=1$.\label{fig:exp3_n1_robust_gains}]{\includegraphics[scale=0.7]{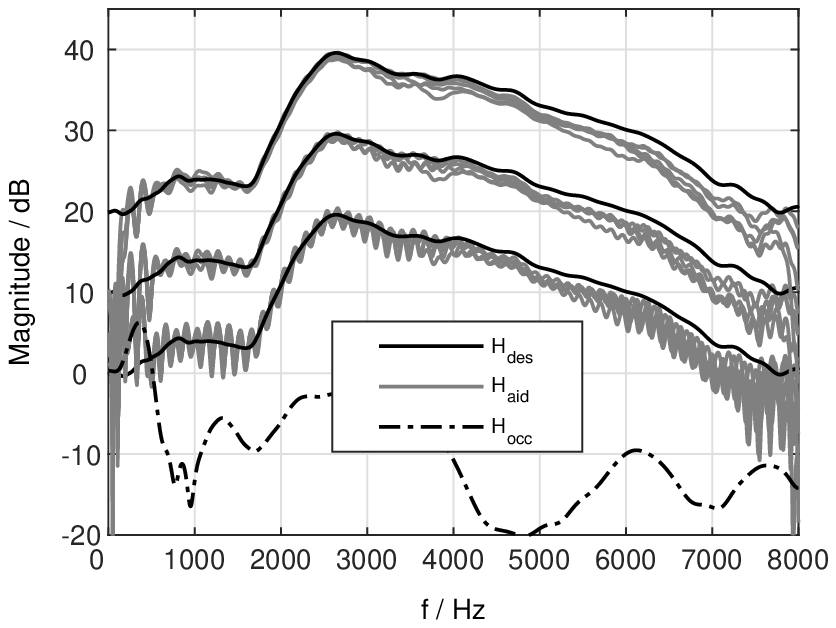}}
	\hfill
	\subfloat[$N=2$.\label{fig:exp3_n2_robust_gains}]{\includegraphics[scale=0.7]{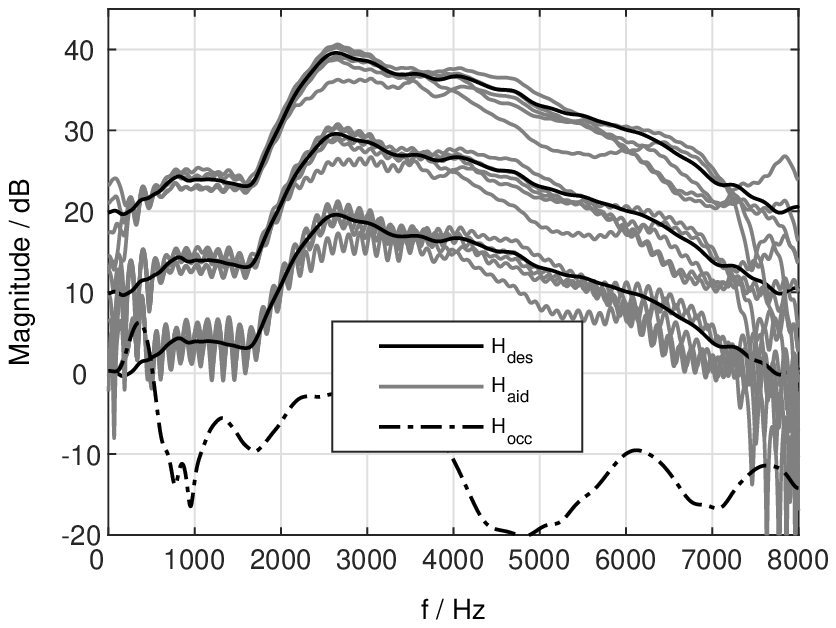}}
	\caption{Magnitude responses of the occluded ear transfer function $H_{occ}(q)$, the desired open ear transfer function $H_{des}(q)$ and the aided ear transfer function $H_{aid}(q)$ for the robust optimization for different forward path gains $G_0$ and (a) $N=1$ and (b) $N=2$ ($\lambda=0.1$, $d_G=96$, $d_H=32$, $\beta=1$). The forward path gains are from bottom to top $G_0 = 0$\,dB, $G_0=10$\,dB, and $G_0=20$\,dB.} \label{fig:exp3_robust_gains}
\end{figure*}

For all considered forward path gains, Figure \ref{fig:exp3_robust_gains} shows the magnitude responses of the aided ear transfer function for $N=1$ and $N=2$ loudspeakers, respectively, as well as the desired open ear transfer function and the occluded ear transfer function. As can be observed, a similar equalization performance is achieved for the different forward path gains. Furthermore, as expected comb-filtering effects are reduced with larger forward path gains due to the reduced impact of the leakage component on the aided ear transfer function (see Section \ref{sec:systemanalysis}). In addition, it can be observed that the effect of the forward path gain is similar for $N=1$ and $N=2$ loudspeakers. In conclusion, these results demonstrate that the proposed approach enables to achieve a good equalization performance for different forward path gains, independent of the number of loudspeakers used without changing the design parameters, i.e., the acausal delay $d_H$, the regularization constant $\lambda$ and the control parameter $\beta$.

\section{Conclusion \label{sec:conclusion}}
In this paper we considered a least-squares-based procedure to design single- and multi-loudspeaker equalization filters for hearing devices aiming at achieving acoustic transparency. We proposed a unified design procedure for both single and multiple loudspeakers to compute the equalization filter by minimizing a least-squares cost function. We showed that for the considered scenario the multi-loudspeaker system exhibits common zeros and propose to exploit the exact knowledge about these common zeros and reformulated the optimization problem accordingly. Since with increasing delay of the hearing device processing comb-filtering artifacts are one of the major limitations to achieve a high quality of the sound at the ear drum, we proposed to reduce the hearing device playback when the leakage signal and the desired signal at the eardrum are of similar magnitude by incorporating a frequency-dependent regularization in the equalization filter design. In order to improve the robustness to unknown acoustic transfer functions, we propose to consider multiple sets of measured ATFs in the design of the equalization filter. Experimental results using measured ATFs from a multi-loudspeaker earpiece show that both using a single loudspeaker as well as multiple loudspeakers a robust equalization can be achieved when considering a robust filter optimization based on multiple sets of measurements, where single-loudspeaker equalization is slightly more robust than multi-loudspeaker equalization. Furthermore, the results show that the proposed frequency-dependent regularization is able to reduce comb-filtering artifacts mainly in the lower frequency regions.



\begin{backmatter}


\section*{Funding}
This work was funded by the Deutsche Forschungsgemeinschaft (DFG, German Research Foundation) – Project ID 352015383 (SFB 1330 A4 and C1), and Project ID 390895286 (EXC 2177/1).

\section*{Competing interests}
The authors declare that they have no competing interests.

\section*{Authors' contributions}
H.S. contributed in developing the main algorithmic idea, deriving the mathematical analysis, performing simulations, analyzing the simulation results, and drafting the article. F.D. contributed in developing the main algorithmic idea, analyzing the simulation results and revising the article. B.K. and S.D. contributed in critically discussing the mathematical analysis, the simulation results and revising the article. All authors read and approved the final manuscript.



\bibliographystyle{bmc-mathphys} 


\end{backmatter}
\end{document}